# Biredox ionic liquids with solid-like redox density in the liquid state for high-energy supercapacitors

Eléonore Mourad[1, 4‡], Laura Coustan[1, 4‡], Pierre Lannelongue[1,4], Dodzi Zigah[2], Ahmad Mehdi[1], André Vioux[1], Stefan A. Freunberger[3], Frédéric Favier[1, 4] and Olivier Fontaine[1, 4*]

[1]Institut Charles Gerhardt Montpellier, UMR 5253, CC 1701, Université Montpellier, Place Eugène Bataillon, 34095 Montpellier Cedex 5, France

[2]Université Bordeaux, ISM, CNRS UMR 5255, F-33400 Talence, France

[3]Institute for Chemistry and Technology of Materials, Graz University of Technology, Stremayrgasse 9, 8010 Graz, Austria

[4]Réseau sur le Stockage Electrochimique de l'énergie (RS2E), FR CNRS

[‡]These authors contributed equally to this work.

* to whom correspondence should be addressed: olivier.fontaine@univ-montp2.fr

**Kinetics of electrochemical reactions are several orders of magnitude slower in solids than in liquids as a result of the much lower ion diffusivity. Yet, the solid state maximizes the density of redox species, which is at least two orders of magnitude lower in liquids because of solubility limitations. With regard to electrochemical energy storage devices, this leads to high-energy batteries with limited power and high-power supercapacitors with a well-known energy deficiency. For such devices the ideal system should endow the liquid state with a density of redox species close to the solid state. Here we report an approach based on biredox ionic liquids to achieve bulk-like redox density at liquid like fast kinetics. The cation and anion of these biredox ILs bear moieties that undergo very fast reversible redox reactions. As a first demonstration of their potential for high-capacity / high-rate charge storage, we used them in redox supercapacitors. These ionic liquids are able to decouple charge storage from ion accessible electrode surface, by storing significant charge in the pores of the electrodes, to minimize self-discharge and leakage current as a result of retaining the redox species in the pores, and to raise working voltage due to their wide electrochemical window.**

Energy storage is undeniably one of the greatest technological and societal challenges of the 21$^{st}$ century driven by the growing demand for renewable but intermittent energy supplies and mobile power sources.[1] Among storage approaches, electrochemical energy storage appears as the most versatile for multi-purpose uses. Ion exchange processes balancing electronic charges are central to electrochemical energy storage. This mechanism includes solid redox active materials in batteries and surface capacitive or pseudocapacitive storage in supercapacitors[2,3,4,5]. However, ion diffusivity is typically around seven orders of magnitude lower in solids than in liquids. This is reflected by the drastically different kinetics of the electrochemical reactions involved in the solid and liquid state[6,7]. Nevertheless, the crucial advantage of the solid state is to provide a considerably higher density of redox active species. Accordingly, solid bulk storage materials lend high energy density, albeit at modest rate[8,9]. In contrast, electrochemical capacitors can deliver outstanding power thanks to the fast kinetics associated to the electrochemical storage mechanisms at the electrode-electrolyte interface. However, the energy density is limited by the surface density of sorbed ions in electrochemical double layer capacitors (EDLC) or of pseudocapacitive material[2,4,10,11,12,13,14,15,16,17].

The topical challenge for supercapacitors is to increase energy without compromising power. One way is to increase the operating voltage by means of ionic liquid (IL) based electrolytes instead of aqueous or molecular non-aqueous electrolytes[18,19,20,21,22]. ILs have the additional advantage to improve safety and allow designing their constituting ions for specific requirements[7]. The other leverage to improve energy is the capacitance. The recent development of highly porous carbon materials has revealed limits to further improvement of EDLC by higher ion-accessible surface area[4,5,12,13,14,15,16,17,22]. Despite enhanced energy densities, pseudocapacitive storage at the electrode material surface is equally surface limited[23,24,25]. This limitation could be overcome in principle by involving redox species dissolved in the electrolyte. This was recently reported for aqueous electrolytes with, e.g., iodide, hydroquinone, $VOSO_4$ or p-phenylenediamine[26,27,28,29,30,31]. These works demonstrated that dissolved redox species reacted with the same fast kinetics as the electrostatic surface storage mechanisms. However, only little additional capacitance was recovered at prohibitively high self-discharge, because of low solubility and high ion diffusivity[32]. Recently, some of us synthesized a new IL with a redox moiety attached to one ion. Electrochemical characterization suggests its efficiency as redox active electrolyte in a Li-ion battery, despite severe electrode balancing and self-discharge issues[33,34,35,36,37]. These results prompted us to turn to ILs with redox active moieties on both ions (making them bulkier) to obtain electrochemical devices with high energy density and limited self-discharge.

Here, we report a biredox IL, where anion and cation are functionalized with anthraquinone (AQ) and 2,2,6,6-tetramethylpiperidinyl-1-oxyl (TEMPO) moieties, respectively, and which demonstrates improved properties when used in model supercapacitors. Tethering a redox group to an ion makes this group ionic in either oxidation state, which increases its solubility in IL media and thus raising the redox density while keeping liquid-state reaction rates. Hence faradaic charge storage becomes decoupled from the ion-accessible electrode surface limitation. This gives access to higher capacity towards that of solid redox materials at the fast redox kinetics of dissolved redox species. The measured capacitance is twice that with non-redox IL electrolyte and was sustained for 2000 cycles without deterioration. The bulky size of the redox ions impeded diffusion through the porous electrodes, which curbed self-discharge to the level of the redox non-active IL electrolyte. This new electrolyte concept opens up new opportunities to develop high-energy supercapacitors and a wide new field in redox materials.

The concept of the biredox IL enhanced capacitor in comparison to the purely capacitive EDLC is illustrated in Fig. 1. The biredox IL comprises a perfluorosulfonate anion bearing anthraquinone (AQ–PFS$^-$) and a methyl imidazolium cation bearing TEMPO (MIm$^+$–TEMPO$^\bullet$). Cells comprise carbon electrode materials and either pure butylmethyl imidazolium bis(trifluoromethylsulfonimide) (BMImTFSI) IL, biredox IL dissolved in BMImTFSI or pure biredox IL (at 60°C) as the electrolyte. Since the here shown biredox IL melts above 9 °C (see differential scanning calorimetry, DSC, in Fig. S6) it could only be used at room temperature as a salt dissolved in non-active BMImTFSI or at 60°C when pure. When the cell is charged with pure BMImTFSI electrolyte cations and anions are drawn into the negative and positive electrode, respectively, and are adsorbed at the carbon surface without undergoing a Faradaic reaction. In the case of the biredox IL containing cell, the same processes apply to the BMIm$^+$ cation and TFSI$^-$ anion. Additionally, as the redox active AQ–PFS$^-$ and MIm$^+$–TEMPO$^\bullet$ ions are electrosorbed at the surface of the carbon electrodes, they undergo fast Faradaic reactions. Their bulky size in combination with the high viscosity of the electrolyte impedes self-discharge as further discussed below. Since either redox species is ionic in either oxidation state the solubility greatly exceeds the typical solubility of neutral redox species. Here we attach the reducible moiety to the anion and the oxidizable one to the cation, which results in bi-anions and bi-cations being generated at anode and cathode, respectively. However, also the opposite combination would be perceivable, which would result in zwitterionic species.

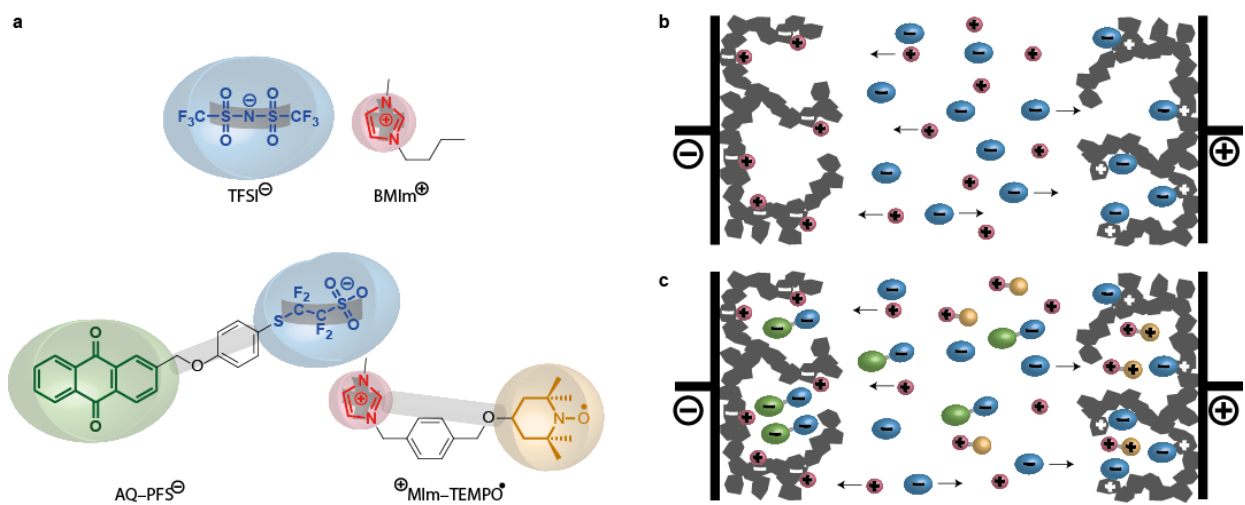

**Figure 1| Comparison of charge storage in EDLC with IL electrolyte and the biredox IL enhanced capacitor. a**, Structure of the here used BMImTFSI IL and the biredox IL comprising a perfluorosulfonate anion bearing anthraquinone (AQ–PFS⁻) and a methyl imidazolium cation bearing TEMPO (MIm⁺–TEMPO•) . **b–c**, Charge storage in a purely capacitive EDLC comprising porous carbon electrodes and an IL electrolyte (**b**) and the here developed concept of capacitors with additional Faradaic processes at the redox active ions of the biredox IL electrolyte (**c**).

We prepared the biredox IL by first separately synthesizing a methyl imidazolium cation bearing a TEMPO moiety, and the lithium salt of a perfluorosulfonate anion bearing an anthraquinone (AQ) moiety and then carrying out the metathetic reaction to obtain the biredox IL, Fig. 2. The first involves a Williamson ether synthesis from 4-hydroxyl-TEMPO and α-α'-dibromo-*p*-xylene followed by a quaternization reaction with 1-methylimidazole to yield MIm⁺–TEMPO• Br⁻ (**1**). The synthesis of the Li perfluorosulfonate bearing an AQ moiety (Li⁺ AQ–PFS⁻, **2**) follows an analogous path via ether synthesis from chloromethyl anthraquinone and the appropriate alkoxide. Finally the two intermediates are combined, yielding the biredox final IL. More experimental details are given in the supplementary information.

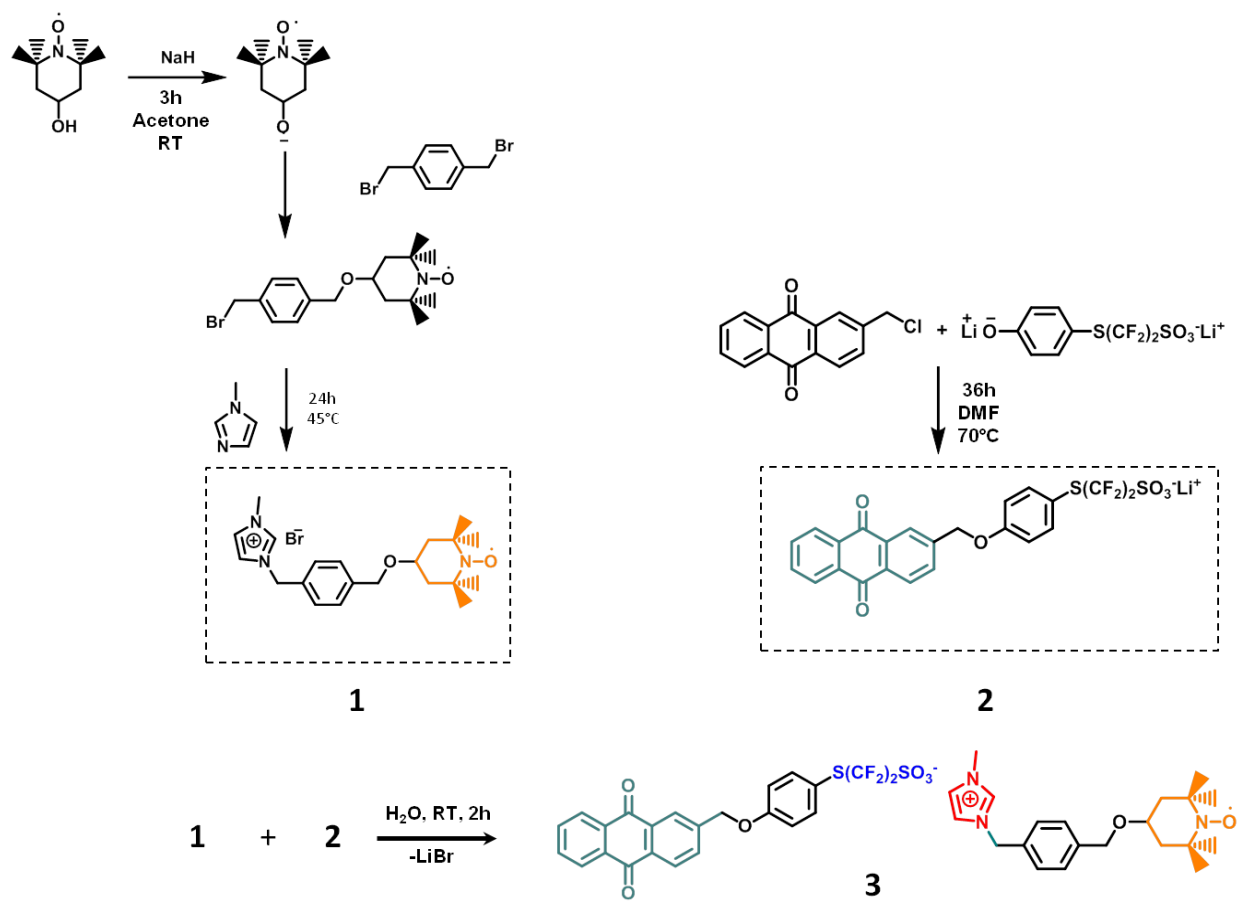

**Figure 2 | Synthesis of the biredox ionic liquid.** The cation and anion bearing AQ and TEMPO are prepared separately before obtaining the biredox IL through metathesis.

IR analysis confirms the presence of the characteristic bands for TEMPO and AQ at around 2900 and 1300 cm$^{-1}$, Fig. S7. The spectrum of **3** does not correspond to a simple addition of the spectra of the constituting precursors **1** and **2** before the metathesis reaction. This suggests that, after ion exchange and removal of LiBr, the resulting moieties experience a different environment than in the initial salts **1** and **2** (with Li$^+$ and Br$^-$ counter ions, respectively). The additional bands in the spectrum of **3** indicate new interactions that likely originate from specific interactions between the sulfonate anion and the imidazolium cation. Biredox IL, precursors and intermediates were characterized by mass spectrometry (Fig. S2, S4, S5), differential scanning calorimetry (DSC) (Fig.S6) and thermogravimetric analysis (TGA)(Fig. S8). NMR (Fig. S3) was used for the characterization of the anion bearing anthraquinone (TEMPO is actually a paramagnetic compound). The water content of the final biredox IL in BMimTFSI electrolyte was measured by Karl Fischer (80 ppm and 100 ppm respectively). These data demonstrate the composition and purity of the prepared biredox IL.

The cyclic voltammograms (CV) of 2mM biredox IL and 0.1 M $TBAPF_6$ solutions in acetonitrile (MeCN) when using a glassy carbon disc electrode are shown in Fig. S9. The peak couple around 0.6 V vs. Ag/AgCl is assigned the oxidation of the TEMPO radical ($MIm^+$–$TEMPO^\bullet$) to the TEMPO cation ($MIm^+$–$TEMPO^+$) and its reverse reaction. The reduction of the AQ–$PFS^-$ moiety to the anthraquinone radical anion ($AQ^{\bullet-}$–$PFS^-$) and di-anion ($AQ^{2-}$–$PFS^-$) is characterized by two peak couples around -0.7 and -0.9 V vs. Ag/AgCl.[38,39] The CV confirms equally reversible electrochemistry of the redox moieties when attached to the ions as in the well-known unsupported substances (see Fig. S16 for the redox reactions). The electrochemical stability window with glassy carbon electrodes is 3.7 V, for the 0.5 M biredox IL solution in BMImTFSI (Fig. S10). In asymmetric PICA/PICA device it still exceeds 2.8 V, demonstrating the advantage of using IL electrolytes to expand the potential window to nearly 3 V.[36] This stands in contrast to previously reported dissolved redox species in supercapacitors like AQ or iodide in aqueous media where the electrochemical stability window is limited to ≈ 1.2 V.[37,38] The dissolution of biredox IL in BMImTFSI up to saturation slightly increased the conductivities (Table S1 and Fig. S11). The conductivity of the pure biredox IL was several orders of magnitude lower and could only be measured above its glass transition temperature. In agreement with the literature, the viscosity of BMimTFSI was measured at 53 cP and its density at 1.4.[39]

**The biredox IL in capacitors**

Symmetric supercapacitors were assembled as detailed in the Supplementary Information (Methods) with either BMImTFSI, 0.5 M biredox IL in BMImTFSI or pure biredox IL as the electrolyte. Electrodes were based on PICA and YP50 activated carbons and reduced graphene oxide (rGO). We have chosen these three carbons with markedly different pore sizes and pore size and distributions to probe surface accessibility for widely varying ion sizes of the BMImTFSI and biredox IL as discussed later.

Cyclic voltammograms of the PICA and rGO cells with BMImTFSI and 0.5M biredox IL in BMImTFSI are compared in Fig.3a and b, where in either case a doubling of the capacitance is observed. With the predominantly nanoporous YP50 carbon any significant change is observed between measurements in these two electrolytes (Fig. S13). The CV curve of an ideal EDLC supercapacitor features a rectangular shape since the capacitance $C$ is constant and independent of the applied voltage at constant voltage scan rate. As expected, the CVs measured with pure BMImTFSI electrolyte (dashed line in Fig. 3a and c, Fig. S13) reveal, particularly with PICA and YP50 electrodes, a quasi-rectangular shape. These shapes are characteristic of an EDLC behavior without or very limited redox contribution. For the rGO electrode, the minor redox peaks below 1.8 V can be assigned to surface groups on the carbon. Capacitances are calculated by integrating the CVs according to the de Levie model for supercapacitors (equation (S1)).[40] Specific capacitances measured at room temperature for PICA, rGO, and YP50 based devices when cycled at 5 mV·s$^{-1}$ as well as self-discharge and leak current are reported in Table S3.

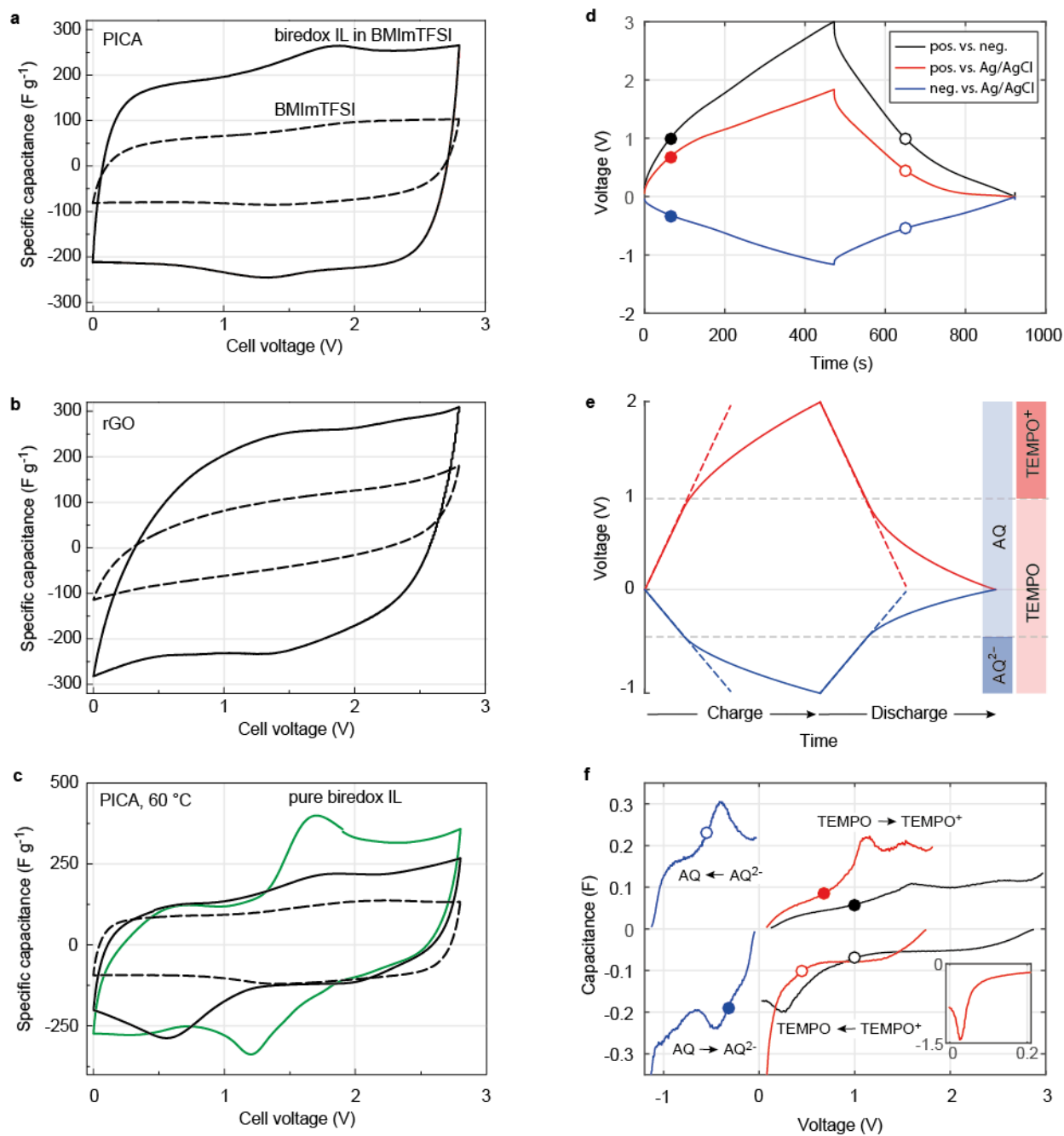

**Figure 3 | Electrochemistry of the biredox IL in the supercapacitors. a**, **b**, Cyclic voltammetry at 5 mV·s$^{-1}$ with 0.5 M biredox IL in BMImTFSI (solid line) and pure BMImTFSI (dashed line), respectively, and electrodes made with PICA (**a**) rGO (**b**). c, CVs at 5 mV·s$^{-1}$ of the PICA cell at 60 °C with 0.5 M biredox IL in BMImTFSI (solid line), pure BMImTFSI (dashed line), and pure biredox IL (green line) **d** – **f**, galvanostatic cycling data in a three-electrode cell with PICA working electrodes and a Ag/AgCl reference electrode and 0.5 M biredox IL in BMImTFSI at room temperature and 1.5 A·g$^{-1}$. Data are given for the full cell (black) and the positive (red) and negative (blue) electrodes versus reference, respectively. The filled and open circles denote the points, where the cell passes 1 V on charge and

discharge, respectively. **e,** Schematic representation of the galvanostatic voltage response of the electrostatic and faradaic processes with the biredox IL and the stability regions for its oxidized/reduced species. The dashed lines illustrate the behavior if only electrostatic processes would prevail.**f,** Capacitances $C = dQ/dU$ versus full cell or single electrode voltage.

Turning to 0.5 M biredox IL in BMImTFSI as electrolyte, the CV curves for PICA and rGO electrodes (solid lines in Fig. 3a and b) display a similar shape as with BMImTFSI but with drastically increased amplitudes (from 100 to 200 F·g$^{-1}$ for PICA activated carbon) over most of the voltage range. Additionally, broad oxidation and reduction peaks appear at intermediate voltages that point out the redox processes of the biredox IL. As shown in Table S3, specific capacitances are roughly doubled in comparison with pure BMImTFSI electrolyte, which demonstrates the potential of the biredox IL to enhance device capacitance and energy density.

Activated carbons have been the most commonly used electrode materials for EDLC over the past years due to their high specific area, low cost, excellent chemical and thermal stability as well as relatively good conductivity and ease of modification. Particularly, the nature and ionic permittivity of nanostructured carbons play are critical to the specific capacitance. Here, we compare PICA with rGO and YP50 carbons. In both electrolytes, CVs of the cells with rGO electrodes (Fig. 3b) appear slightly more distorted than those with PICA (Fig. 3a) or YP50 (Fig. S13) electrodes. These distortions are characteristic of Ohmic contributions to the behavior of the device. Careful analysis of the measurements confirms that this Ohmic behavior can neither be assigned to the electrolyte nor to the assembling of the Swagelok cell. Moreover, this Ohmic distortion persists for both electrolytes. Unlike the other carbons, the performance of the YP50 electrodes does not change noticeably when switching the electrolyte as the specific capacitance remains at ≈ 110 F·g$^{-1}$ in either case (Fig. S13). Since the capacitance of carbon based SCs is related to the ion accessible surface area, this behavior can be explained by the markedly different pore size distribution of the respective materials. The pore size distributions of the selected carbons as derived from the N$_2$ sorption isotherms are given in Fig. S12. Reduced graphene oxide (rGO) has a wide pore size distribution without prominent discrete pore sizes while the activated PICA and YP50 carbons are microporous with a significant amount of pores centered at about 1 nm. Figure S12c highlights the key difference between PICA and YP50 in terms of mesoporosity. YP50 is nearly a purely microporous material as most of the pore volume (66% of the pore volume) stems from micropores with an average diameter around 0.9 nm. Its mesoporosity accounts for 0.5% of the pore volume only. In contrast, rGO offers an open surface with theoretically unrestricted access for the electrolyte species, while the porous structure of PICA is more complex with micropores below 1 nm diameter (25% of the pore volume) together with mesopores ranging from 2 to 5 nm in diameter. The volume associated with mesopores represents 49% of the total pore volume. We assume that the difference between YP50 and the other tested carbons stems from the different accessibility of the available pore space for the electrolyte species. In YP50 the biredox molecules cannot enter the micropores, while the mesopores of PICA and the open surface of rGO are accessible. In contrast, BMImTFSI in contrast appears to access much smaller pores than accessible for the biredox species. Since PICA based electrodes show the best compromise between conductivity and pore size distribution the further characterizations use this carbon.

The enhancement of the PICA specific capacitance scales with the concentration of biredox IL in BMImTFSI in the range from 0 to 0.5 M, Fig. S16 and Table S4. At room temperature the viscosity of the saturated solution (≈0.62 M) becomes prohibitively high, which causes the capacitance to drop in comparison with the 0.5 M solution. Further improvement can however be achieved with pure redox IL as the electrolyte, but requires an elevated temperature. Specific capacitance of a PICA based cell at 60 °C are compared when operated with BMImTFSI, 0.5 M biredox IL in BMImTFSI, and pure biredox IL (Fig. 3c). In the case of the redox active electrolytes redox peaks are clearly visible. Despite the redox features displayed by the CVs, especially in pure biredox IL, average capacitances have been calculated by CV integration. The values are 99, 204, and 370 $F \cdot g^{-1}$ for 0, 0.5 M and pure biredox IL, respectively. The nearly unchanged values between room temperature and 60 °C for the first two electrolytes demonstrate that the increase is associated with redox concentration. The remarkable value for pure biredox IL compares favorably well with the capacitance of pseudocapacitive solid materials such as $MnO_2$. The enhancement in comparison to the pure BMImTFSI can hence definitively be assigned to the redox activity of the biredox IL.[40,41,42] The nearly unchanged capacitances from room temperature to 60 °C for the first two electrolytes demonstrate that this increase cannot be associated to any other characteristic changes of the electrolyte, including ionic conductivity enhancement.

The data with the pure molten substance demonstrate the ultimate possibility of that the biredox IL permits. The density of redox groups equals that of the bulk substance, yet the kinetics of its redox reactions is similar to that of a redox molecule in solution. Therefore, the biredox IL unites the advantage of high redox density that is otherwise only found in solids and the fast kinetics of solution species/systems. Considering the porosity of the PICA electrode (≈64 %) and the concentration of the pure biredox IL (3.4 M), half of the redox groups in the pores have reacted and contributed to measured capacitance. Therefore, biredox ILs, with the extreme case of the pure melt, allow achieving bulk density of redox species with fast liquid like charge transfer kinetics.

The CV curves of cells with biredox IL electrolytes show a pair of more or less pronounced peaks at intermediate cell voltages. These can be assigned to a convolution of the redox reactions at anode ($AQ/AQ^{2-}$) and cathode ($TEMPO/TEMPO^+$). They are overlapped with electrostatic contributions from the electrosorption of anions and cations (either bearing redox species or not). The corresponding redox reactions concerned at negative and positive electrodes and therefore the perfluorosulfonate anion bearing anthraquinone ($AQ–PFS^-$) and the methyl imidazolium cation bearing TEMPO ($MIm^+–TEMPO^{\bullet}$) are given in Fig. S18. The reactions are to be understood in addition to the non-faradaic sorption of cations/anions at the negative and positive electrode, respectively. When the SC is charged, the $AQ–PFS^-$ is adsorbed at the negative electrode and is reduced as $AQ^-–PFS^-$ or further to $AQ^{2-}–PFS^-$. Concurrently, $MIm^+–TEMPO^{\bullet}$ is adsorbed at the positive electrode and oxidized as $MIm^+–TEMPO^+$. Since $AQ–PFS^-$ and $MIm^+–TEMPO^{\bullet}$ are in a ratio of 1:1, the contribution of AQ to the capacitance is theoretically twice that of the TEMPO. To clarify their role in the charge/discharge mechanism, galvanostatic cycling data were collected in a three-electrode cell with symmetric PICA electrodes, an Ag/AgCl reference electrode and 0.5 M biredox IL in BMImTFSI as the electrolyte (Fig. 3d to f). The black curve corresponds to the response of the full cell, measured between negative (blue curve) and positive (red curve) electrodes versus the reference. The cell voltage profile during discharge and charge, Fig. 3d, can be divided into two main parts, a steeper section at the beginning of charge or discharge that represents mostly capacitive processes and a flatter section afterwards corresponding to additional redox reactions. This is schematically shown in Fig. 3e together with the stability regions of the reduced/oxidized redox species. To make this assignment clearer, capacitances were calculated for the full cell as well as for each electrode via $C = dQ/dU$ with $Q$ and $U$ being charge and voltage, respectively. The result is plotted in Fig. 3f versus cell voltage or single electrode voltage, respectively. The filled and open circles denote the point where the full cell passes 1 V and aid assigning single electrode processes to the full cell curve.

The plot of the full cell response versus voltage, Fig. 3f, shows a rectangular shape with additional broad anodic and cathodic peaks akin to the CV data (Fig. 3 a to c). A close inspection of the single electrode curves at the onset of the anodic peak (black dot, 1 V) shows that the oxidation of the $MIm^+$–$TEMPO^\bullet$ just starts (red dot) while the $AQ$–$PFS^-$ reduction is already ongoing (blue dot). Similarly, on discharge the onset of the cathodic peak (black circle, 1V) corresponds to that of the $MIm^+$–$TEMPO^+$ reduction (red circle) while the $AQ^{2-}$–$PFS^-$ oxidation has already started (blue circle). The assignment of the redox processes in the single electrodes is supported by the similar potentials for $MIm^+$–$TEMPO^\bullet$/$MIm^+$–$TEMPO^+$ and $AQ$–$PFS^-$/$AQ^{2-}$–$PFS^-$ versus the same reference electrode when measured either on a glassy carbon disc or with a PICA electrode (Fig. S19). Overall, the galvanostatic curves in Fig 3e confirm the main characteristics of a redox capacitive behavior with redox reactions involved over a wide range of the voltage window.[46-47]

**Self-discharge and leak current**

Self-discharge and leakage currents are the main concerns when using redox active electrolytes. It is usually identified by a decay of the open circuit voltage with time after charging. In unfavorable cases this voltage drop can occur within several minutes, quickly leading to an unusable device. In EDLC, this phenomenon is originating from a redistribution of charge among the capacitive elements. It is usually rather limited for EDLC operated in classical $NEt_4BF_4$/acetonitrile or IL-based electrolytes as the ions compensating the electrode charges are "trapped" in the double layer at the electrode-electrolyte interface. In the case of dissolved redox active species, they are eventually diffusing and the migration to the opposite electrode can cause their re-reduction or oxidation, respectively, leading to a discharge current by this shuttle effect and a loss of the stored charge. However, in the case of the biredox IL presented here the reduced/oxidized species are doubly charged ions, which may to some extent remain electrosorbed at the charged electrode and may therefore reduce free diffusion and self-discharge. Self-discharge experiments were done by measuring the time within which the voltage drops from the fully charged state to half of the voltage, *i.e.*, from 2.8 to 1.4 V in the present case. Leakage current was measured by holding the potential at full charge.

Before self-discharge and leakage current measurements, cells with the various carbon materials were assembled as described and initially charged with a current of 1 mA·g$^{-1}$ to 2.8 V and held at this voltage for 2 hours. The results are listed in Table S3 and Fig. 4. With values between 30 and 40 µA·F$^{-1}$·V$^{-1}$, the biredox IL cells show a 2 to 3 times lower leak current than the cells with BMImTFSI electrolyte with ≈ 100 µA·F$^{-1}$·V$^{-1}$ leak current. After full charge, the cells were allowed to undergo self-discharge for 50 hours while monitoring the open circuit voltage decay.

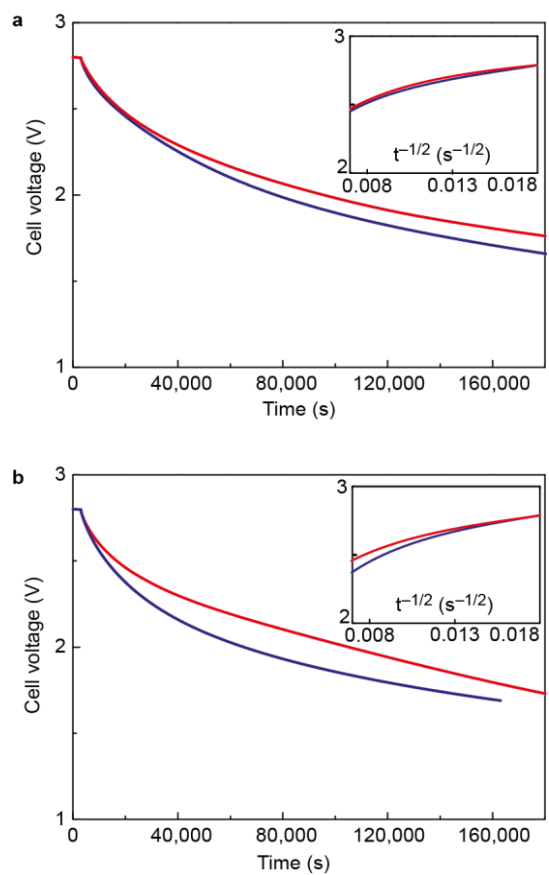

**Figure 4| Self discharge of supercapacitors. a,b,** Self-discharge current test for the devices based on PICA (**a**) and rGO (**b**) with BMImTFSI (blue line) or 0.5 M biredox IL in BMImTFSI (red line) at room temperature. Cells were first charged at 1mA·g$^{-1}$ to 2.8 V. Calculated self-discharge and leakage currents are given in Table S3. Corresponding data for the YP50 device are given in Fig. S13b.

Similar voltage profiles were obtained for all tested devices as shown in Fig. 4a, b and S13b. During the first 6 h all SCs show relatively rapid voltage decay down to about 2.5 V. This is followed by a slower decay to reach a 1.7 – 1.8 V plateau. Corresponding self-discharge currents measured by coulometry are listed in Table S3. Devices based on pure BMImTFSI yield self-discharge currents of about 4 µA·F$^{-1}$. When operated with biredox electrolyte the leakage currents are 2 to 3 times smaller. In either case, the observed self-discharge currents in the range of a few µA·F$^{-1}$ are fairly small. Self-discharge only weakly depends on the electrode nature as behaviors are very similar for both PICA and rGO based devices. The slightly lower figures for the PICA cell suggests the "entrapment" of the biredox species within the porous structure to impede self-discharge, albeit some charge shuttling is expected to still take place[43]. The lower self-discharge in the biredox electrolyte may also be aided by the electrosorption of the multi charged ion as well as the higher viscosity of the electrolyte. As self-discharge and current leakage are associated to the undesired diffusion of charge carriers, discrepancies can usually be highlighted by comparing the device voltage decay as a function of $t^{-1/2}$ (insert Fig. 4). For a diffusion controlled behavior, a linear voltage decays is expected. In both BMImTFSI and biredox electrolytes as well as for PICA and rGO based devices, decays are non-linear versus $t^{-1/2}$ (Fig. 4 a, b and S13b). These deviations are assigned to anomalous diffusion within the porous electrodes and the diffusion layer thickness is not directly proportional to the square root of time but may be fitted to an exponential factor smaller than 0.5.

**Power capability**

The power capability of the series of SCs described above was evaluated by measuring the capacitance as a function of rate. Figure S22 shows the specific and relative capacitances measured by cyclic voltammetry for each cell at scan rates ranging from 5 to 200 mV·s$^{-1}$. All devices, PICA (square), rGO (triangle) and YP50 (diamond) show an associated decay of capacitance when the scan rate is raised. This observed loss of power is roughly the same for the biredox (red solid line) and BMImTFSI (blue dashed line) electrolytes and cannot be assigned to the presence of biredox moieties.

For a better understanding of the effect of biredox species in the electrolyte, the energy and power densities of various PICA based symmetric cells when operated with different electrolytes are compared in the Ragone plot in Fig. 5. Details for the calculations of energy and power densities are given in SI. Base case electrolytes are 0.5 M NEt$_4$BF$_4$ in acetonitrile (ACN) and pure BMImTFSI. The first excels with power density whereas the latter allows for higher energy. Diluting BMImTFSI with acetonitrile allows for higher power density at high rate at unchanged peak energy density, which can be ascribed to lowered viscosity. Addition of 0.5 M biredox IL to BMImTFSI doubles the energy at low rate to 70 Wh·kg$^{-1}$ and keeps energy and power density well beyond pure BMImTFSI at all rates. As such, this electrolyte appears the best compromise in terms of power and energy densities over the whole range of rates. As expected, the biredox electrolyte can, however, not compete for peak power with the widely used NEt$_4$BF$_4$ in acetonitrile electrolyte (solid triangle). A slight gain in energy density over the latter is achieved with a more complex mixture of 0.5 M biredox IL and 0.5 M NEt$_4$BF$_4$ in acetonitrile (black pentagon). This gain in specific energy is however at the expense of specific power. The biredox IL in BMImTFSI (red squares) shows at all rates higher values than in ACN (black pentagon) even though higher peak power could be expected from the lower viscosity of the ACN electrolyte. CVs of the MIm$^+$–TEMPO/MIm$^+$–TEMPO$^+$ couple in both electrolytes are shown in Fig S26 and point at the limiting processes. The kinetics in BMImTFSI is faster as indicated by the small peak splitting but the peak current is restricted by the lower diffusivity. The higher power in the BMImTFSI may therefore be ascribed to the faster kinetics of redox moieties close to the electrode surface, which do not have to diffuse far to react and which emphasizes that towards solid like redox density in the liquid state can be accessed at high rates. The faster kinetics in the IL system may in part be due to the higher supporting electrolyte concentration.

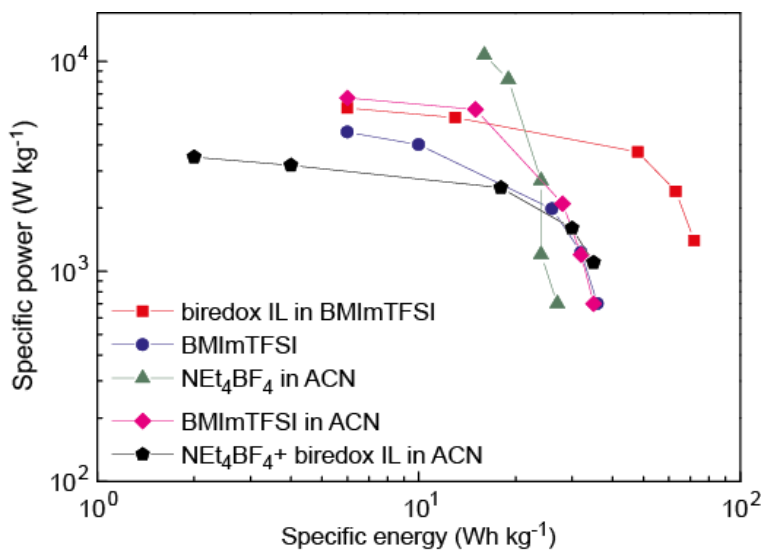

**Figure 5| Ragone plot of supercapacitors using various electrolytes.** 0.5 M biredox IL in BMImTFSI is compared to other electrolytes in symmetric PICA based supercapacitors: 0.5 M NEt$_4$BF$_4$ in acetonitrile (ACN), 0.5 M BMImTFSI in acetonitrile and 0.5 M NEt$_4$BF$_4$ + 0.5 M biredox IL in acetonitrile. Energy and power densities are relative to the weight of PICA carbon in both electrodes. Calculation details are given in ESI.

This power behavior is in agreement with similar ionic conductivities of BMImTFSI and 0.5 M biredox IL in BMImTFSI, and that the charge transfer of the involved redox reactions is fast, Fig. S11 and S14. In a conventional EDLC device, the electrostatic nature of the charge-discharge processes leads to a specific power, which can be obtained by using equation (S4). The maximum operating voltage is determined by the thermodynamic stability of the electrolyte and may be further influenced by any electrocatalytic activity of the electrode materials that could promote electrolyte decomposition. The equivalent series resistance (ESR) is not only composed of the solution resistance but also of the intrinsic resistance of the materials, contact and current collector, diffusion resistance of ions in the electrode materials and through the separator. Therefore, to achieve attractive performance, it is critical to minimize all these resistances. Additionally, charge transfer Faradaic reactions come into play in redox capacitive devices. These resistances together may be characterized by the characteristic time constant of the device, i.e., the time to charge and discharge the device. It may be obtained by fitting the cyclic voltammogram (Fig. S23) using equations (S2) and (S3).[48]

**Cycling stability**

Figure 6 shows the specific capacitance for each device over 2000 galvanostatic cycles at a rate of 1.5 A·g$^{-1}$ using pure BMimTFSI (blue markers) and the biredox IL (red markers) as electrolytes. The galvanostatic measurements lead to the same observations as cyclic voltammetry and the use of the biredox electrolyte doubles the specific capacitance, except for YP50 sample for which the specific capacitance remains the same as with pure BMimTFSI (Fig. S15). The long-term cycling behavior of biredox IL using PICA electrodes show a nearly constant specific capacitance for 2000 cycles after an initial decrease from 180 to 155 F.g$^{-1}$ during the first 200 cycles. Some fluctuations are also visible for rGO with biredox IL and YP50 with both electrolytes. Previous works about hydroquinone used as dissolved electroactive species in an aqueous electrolyte have shown that the cycling stability depended on the carbon nature.[49] In the worse cases, after 2000 cycles, the specific capacitance dropped by up to 35%. This serious fading was assigned to a limited reversibility in the reduction/oxidation of hydroquinone within the applied voltage window impacting the Coulombic efficiency.

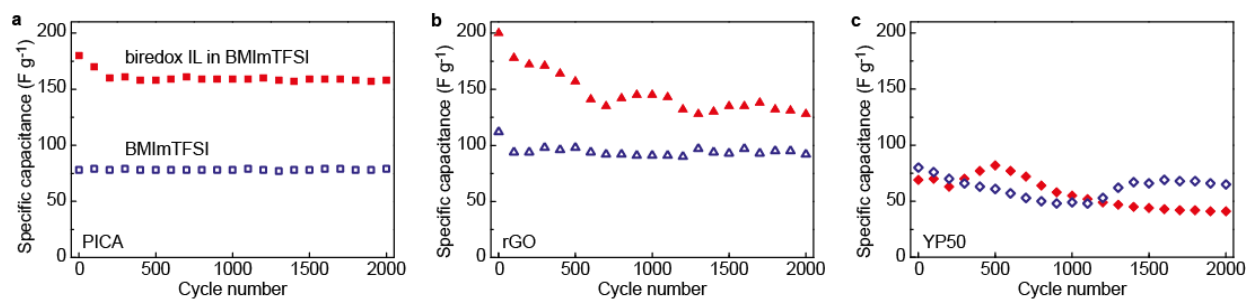

**Figure 6| Specific capacitance retention upon cycling in biredox IL electrolyte.** Data with 0.5 M biredox IL in BMImTFSI as the electrolyte are shown in red with filled markers and data with pure BMImTFSI electrolyte are shown in blue with and open markers. The symmetric devices are based on PICA (a), rGO (b) and YP50 (c) and were cycled at 1.5 A·g$^{-1}$ (based on the mass of a single electrode).

To explain the fading of capacitance in the various cells, the Coulombic efficiency vs resistance and capacitance vs resistance are reported in Fig S14. For the PICA-based cell, the specific capacitance fading during the first 200 cycles is assigned to an increase of the resistance that induces a loss in Coulombic efficiency. This resistance increase is originating either from some irreversible adsorption of anthraquinone at the PICA surface because of its inner-sphere behavior[44,45] and to the protonation of the anthraquinone group because of acidic groups present at the carbon surface.[46] For the rGO –based cell, the Coulombic efficiency remains upon cycling while the capacitance fading perfectly correlates the cell resistance increase. The mechanical instability of rGO, tending to progressively restack upon successive ion intercalation, accounts for the observed resistance increase upon cycling.[47] For the YP50 -based cell, Biredox moieties cannot access the porosity and both resistance and Coulombic efficiency remain stable as only BMImTFSI is involved in the charge storage.

In conclusion we demonstrate a biredox ionic liquid, where anion and cation are functionalized with anthraquinone (AQ) and 2,2,6,6-tetramethylpiperidinyl-1-oxyl (TEMPO) moieties, respectively, and apply it to model supercapacitor electrode materials. Tethering a redox group to an ion makes it ionic in either oxidation state, which allows boosting the density of the redox groups in the liquid state towards the bulk density of redox active solids. This way, high capacity towards that of bulk redox materials and fast redox kinetics concurrent with dissolved redox speciescan be achieved. In contrast to purely electrostatic EDLC and redox capacitive storage using surface immobilized redox species, in such a type of hybrid device the redox capacitive charge storage can be decoupled from the limitation of ion accessible electrode surface and extends to the bulk electrolyte within the pores which properties are expanded beyond mere solvation of species and ion transport. First, we have described the synthesis of the biredox IL along a generally applicable synthetic route. Second, we have assessed the characteristics of the biredox IL by using it in supercapacitors made of three types of large surface area carbons with markedly different pore size distribution. The specific energy of supercapacitors based on activated carbon and graphene oxide electrodes could be doubled with the biredox ionic liquid electrolyte in comparison to the same cell without redox active electrolyte. At the same time the power capability and cycling stability remained preserved. The enhanced capacitance is sustained for 2000 cycles without deterioration. The bulky size of the redox active ions curbs the diffusion through the porous carbon electrodes, which slows down self-discharge to the level of the redox non-active IL electrolyte. Taken together such biredox ionic liquids enable combining an array of properties: accessing towards bulk like redox density at liquid like fast kinetics, surmounting the solubility limit of conventional redox species towards bulk redox density in the liquid state, and when applied to supercapacitors decoupling electrostatic and redox capacitive charge storage of the ion accessible surface area, retaining the redox species in the electrodes to minimize self-discharge and leak current, and raising working voltage and safety in comparison to redox active aqueous electrolytes and acetonitrile based electrolytes, respectively, due to the use of IL electrolyte. The use in supercapacitors represents a first demonstration of the unique possibilities of biredox ionic liquids, which can more generally enable high-capacity / high-rate charge storage. The results represent therefore a solution for the main hurdle of high-energy supercapacitors and open up a wide new field in redox materials and their applications.

**Acknowledgements**

S.A.F. is indebted to la Chaire Total de la foundation Balard for the position of an invited professor at the Institute Charles Gerhardt, Montpellier, France as well as the Austrian Federal Ministry of Economy, Family and Youth and the Austrian National Foundation for Research, Technology and Development and funding from the European Research Council (ERC) under the European Union's Horizon 2020 research and innovation programme (grant agreement No 636069).

**Author Contributions**

E.M. and L.C. contributed equally to this work and carried out the experiments. O.F., F.F., and S.A.F. conceived and designed the experiments, directed the project and analyzed the results. F.F, S.A.F. and O.F. co-wrote the manuscript. A.V. and A.M had helped on synthesis of anionic species. All authors contributed to the discussion and interpretation of the results.

**Competing Financial Interests**

The authors declare no competing financial interests

**Figure Legends, and Tables**

**Figure 1| Comparison of charge storage in EDLC with IL electrolyte and the biredox IL enhanced pseudocapacitor. a**, Structure of the herein used BMImTFSI IL and the biredox IL comprising a perfluorosulfonate anion bearing anthraquinone (AQ–PFS$^-$) and a methyl imidazolium cation bearing TEMPO (MIm$^+$–TEMPO$^\bullet$) . **b–c**, Charge storage in a purely capacitive EDLC comprising porous carbon electrodes and an IL electrolyte (**b**) and the herein developed concept of pseudocapacitors with additional Faradaic processes at the redox active ions of the biredox IL electrolyte (**c**).

**Figure 2 | Synthesis of the biredox ionic liquid.** The cation and anion bearing AQ and TEMPO are prepared separately before obtaining the biredox IL through methathesis.

**Figure 3| Electrochemistry of the biredox IL in the supercapacitors.a**, **b**, Cyclic voltammetry at 5 mV·s$^{-1}$ with 0.5 M biredox IL in BMImTFSI (solid line) and pure BMImTFSI (dashed line), respectively, and electrodes made with PICA (**a**) orrGO(**b**). c, CVs at 5 mV·s$^{-1}$of the PICA cell at 60 °C with0.5 M biredox IL in BMImTFSI (solid line), pure BMImTFSI (dashed line), and pure biredox IL (green line)  **d – f**, galvanostatic cycling data in a three-electrode cell with PICA working electrodes and a Ag/AgCl reference electrode and 0.5 M biredox IL in BMImTFSI at room temperature and 1.5 A·g$^{-1}$. Data are given for the full cell (black) and the positive (red) and negative (blue) electrodes versus reference, respectively. The filled and open circles denote the points, where the cell passes 1 V on charge and discharge, respectively. **e,** Schematic representation of the galvanostatic voltage response of the electrostatic and faradaic processes with the biredox IL and the stability regions for its oxidized/reduced species. The dashed lines illustrate the behavior if only electrostatic processes would prevail. **f**, Capacitances $C$ = d$Q$/d$U$ versus  full cell or single electrode voltage.

**Figure 4| Self discharge of supercapacitors.a**,**b,** Self-discharge current test for the devices based on PICA (**a**) andrGO (**b**) with BMImTFSI (blue line) or 0.5 M biredox IL in BMImTFSI (red line) at room temperature. Cells were first charged at 1 mA·g$^{-1}$ to 2.8 V. Calculated self-discharge and leakage currents are given in Table S3. Corresponding data for the YP50 device are given in Fig. S13b.

**Figure 5 | Ragone plot of supercapacitors using various electrolytes.** 0.5 M biredox IL in BMImTFSI is compared to other electrolytes in symmetric PICA based supercapacitors: 0.5 M NEt$_4$BF$_4$ in acetonitrile (ACN), 0.5 M BMImTFSI in acetonitrile and 0.5 M NEt$_4$BF$_4$ + 0.5 M biredox IL in acetonitrile. Energy and power densities are relative to the weight of PICA carbon in both electrodes. Calculation details are given in ESI.

**Figure 6 | Specific capacitance retention upon cycling in biredox IL electrolyte.** Data with 0.5 M biredox IL in BMImTFSI as the electrolyte are shown in red with filled markers and data with pure BMImTFSI electrolyte are shown in blue with and open markers. The symmetric devices are based on PICA (a), rGO (b) and YP50 (c) and were cycled at 1.5 A·g$^{-1}$ (based on the mass of a single electrode)

**Methods**

**General synthesis of biredox ionic liquid:** 2-Chloromethylanthraquinone and 4-hydroxy-TEMPO, were supplied by Sigma-Aldrich Reagent Co., Germany. Lithium bis(trifluoromethylsulfonyl)imide (LiTFSI) was obtained from solvionic, France. 2-(4-Oxydophenylsulfanyl)-1,1,2,2-tetra-fluoroethansulfonate lithium was obtained from ERAS Lab, Grenoble, France. Solvents used during the synthesis were purified to ≥ 99.9 %.

2-(4-Oxydophenylsulfanyl)-1,1,2,2-tetra-fluoroethansulfonate lithium salt (18 mmol, 1 eq) was dissolved in anhydrous dimethylformamide (DMF) and dried by azeotropic distillation using a Dean-Stark receiver (azeotrope toluene-water). Then the 2-chloromethyl anthraquinone (20 mmol, 1.1 eq) was added. The resulting mixture was stirred under argon atmosphere at 80 °C for 24 h. The product was precipitated by addition of dichloromethane (20 mL) and washed with the same solvent (3 x 15 mL) until obtaining **2** as a yellow solid (63 % yield).

Hydroxy-TEMPO (2.9 mmol) was dissolved in a minimum volume of anhydrous acetone. Sodium hydride was added in small fractions (5 mmol, 1.5 eq) upon which hydrogen evolution was observed. Afterwards, the mixture was stirred for 10 minutes at room temperature and α-α'-dibromo-p-xylene (4.3 mmol, 1.5 eq) was added. The resulting mixture was stirred at room temperature for 3 hours. An orange precipitate was obtained and the acetone was removed under vacuum and the resulting orange solid was poured in distilled water (15 mL). The aqueous phase was washed with dichloromethane (3 × 15 mL). The organic layer was dried over magnesium sulfate and the solvent was removed under vacuum. The raw product was purified by flash chromatography ((90/10) cyclohexane/acetone) and the expected product obtained with 65% yield. The bromide derivative was mixed with1-methylimidazole (1 eq) in methanol. The reaction mixture was then heated at 45 °C overnight. After evaporating the solvent **1** was obtained as an orange viscous liquid. 200 mg (1.2 eq) of **2** and 150 mg (1 eq) of **1** were separately dissolved in a minimum volume of water. They were mixed in a single flask and the mixture was placed under stirring for 2 hours at room temperature. The biredox IL was extracted with dichloromethane. The organic layer was washed with distilled water and then dried over magnesium sulfate. The solvent was removed using a rotary evaporator. The biredox IL was obtained as a yellow compound with 80% yield.

**Electrode preparation:** Carbon powders (PICA, YP50, or rGO), conductive carbon black and PTFE (from 60 wt% PTFE dispersion in water, Sigma Aldrich) were manually mixed in a 75/15/10 ratio in acetone until a homogeneous slurry was obtained. This slurry was spread on a glass plate and rolled several times to obtain films with a thickness of about 150 μm. Film discs with a loading between 10 to 20 mg·cm$^{-2}$ were cut and pressed with 10 tons for 5 seconds onto a nickel foam that is used as current collector in a Swagelok-type two electrodes cell. Before electrochemical characterization, the electrodes were dried overnight under vacuum at room temperature and then immersed, under vacuum, into the IL-based electrolyte for 2 h.

**Physical characterization:** NMR Analysis were performed using a Bruker 300 NMR spectrometer. Different deuterated solvents were used according to the solubility of each compound at room temperature. Chemical shift values are given relative to TMS. Fourier transform infrared (FTIR) spectra were recorded in the 4000-400 cm$^{-1}$ range using 32 scans at a nominal resolution of 4 cm$^{-1}$ by means of aBruker AVATAR 320 FTIR spectrometer equipped with an ATR unit. TGA was carried out using a NETSCH 409 PC in air. Compounds were heated to 1000 °C at 10 °C·min$^{-1}$. DSC measurements were carried out on a NETSCH DSC 204-F1 apparatus. DSC thermograms were recorded on raising the temperature from -120 to 150 °C at a heating rate of 10 °C·min$^{-1}$ under nitrogen atmosphere. Time-of-flight mass spectrometry analysis was carried out on a Synapt G2 -S mass spectrometer (Waters) equipped with an ESI source. Mass spectra were recorded in the positive ion mode between 100 and 1500 Da. The capillary voltage was 1000 V and the cone voltage 30 V. The temperature of the ion source and desolvation were 120 °C and 250 °C, respectively. X-Ray Diffraction (XRD) measurements were performed on a Philips X'Pert diffractometer using a CuKα radiation (λ= 1.5405 Ångstrom). Surface characteristics of carbon powders were evaluated by nitrogen sorption isotherms measured at 77 K with a Micromeritics ASAP 2010 equipment. The pore size distributions were calculated by using the non-local density functional theory DFT method from adsorption isotherms. Raman spectra of powders were recorded by using a He/Ne laser (λ= 633 nm).

**Electrochemical characterization:** Measurements were carried out using a Biologic VMP3 potentiostat or a Princeton Applied Research bipotentiostat. The reference electrode, Ag/AgCl in 0.1 M

TBAClO$_4$ in acetonitrile in a separate compartment separated from the exterior with a dense Li-conducting ceramic, was directly immersed in the electrochemical cell. The working electrode was a 2 mm diameter glassy carbon (GC) disc electrode. The counter electrode was a platinium rod. The working GC electrode was polished with 0.05 µm alumina powder followed by washing with water and acetone before each cyclic voltammogram. The test solutions were de-aerated by flowing a stream of N$_2$ gas (N5.0) through the solution for at least 10 min. Symmetric supercapacitors were assembled using PICA, YP50 activated carbons and rGO electrodes. Electrolytes were either 3-butyl-1-methyl imidazolium bis(trifluoromethylsulfonyl)imide (BMImTFSI) IL or mixtures of biredox ILs and BMImTFSI. Open circuit voltage was measured to be about 30 mV with little change for the different electrode material in the series. Leakage and self-discharge currents have been measured after charging the cell at 1 mA from OCV to 2.8 V, holding the voltage for 2 hours followed by an OCV period. Leakage current has been measured by chronoamperometry during the potential holding step. Self-discharge current have been determined by measuring the time passed for the OCV to drop to 1.4 V and calculated with respect to the initially stored charge.